\documentclass[10pt,a4paper]{article} 
\title{Simpler $O(1)$ Query Algorithm for Level Ancestors}
\usepackage{amssymb}
\usepackage[title]{appendix}
\usepackage[bookmarksopen=true]{hyperref}
\usepackage{bookmark}
\newtheorem{lemma}{Lemma}

\author{Sanjeev Saxena\thanks{E-mail: ssax@iitk.ac.in}\\
Dept. of Computer Science and
Engineering,\\ Indian Institute of Technology,\\
Kanpur, INDIA-208 016}

\date{\today}
\begin{document}
\maketitle

\subsection*{\centering{Abstract}}

This note describes a very simple $O(1)$ query time algorithm for
finding level ancestors. This is basically a serial
(re)-implementation of the parallel algorithm of Berkman and Vishkin
(O.Berkman and U.Vishkin, Finding level-ancestors in trees, JCSS, 48,
214--230, 1994).

Although the basic algorithm has preprocessing time of $O(n \log n)$,
by having additional levels or using table lookup, the preprocessing time can be reduced to almost linear or linear.

The table lookup algorithm can be built in $O(1)$ parallel time with $n$ processors and can also be used to simplify the parallel algorithm of Berkman and Vishkin and make it optimal.

\noindent{\textbf{Keywords:}} Level Ancestors; Rooted Trees;
Algorithms; Graphs; Euler Traversal; Parallel Algorithms

\section{Introduction}

In the level ancestor problem, we are given a rooted tree, which is to
be preprocessed to answer queries of the type:
find the $k$ th ancestor of a node $v$ (here, both $k$ and $v$ are
query parameters).

Several sequential and parallel algorithms are known for this problem
\cite{BV,ABA,BC,D,AH,MM}. The level ancestor algorithm of Bender and
Farach-Colton\cite{BC} is conceptually simple and is usually used in
teaching. Their ``simple algorithm'' \cite[Theorem~8]{BC} takes
$O(n\log n)$ preprocessing time and can answer queries in $O(1)$ time.
The algorithm uses long-path decomposition, ladders and jump pointers.
Their algorithm can answer queries using two table lookups. 
Macro-micro algorithm  \cite[Section 4]{BC}, can be used to reduce preprocessing time and space from $O(n \log n)$ to $O(n)$. The macro-micro algorithm is conceptually simple; however, as per one implementation \cite{MCPP},  the algorithm has ``significant implementation complexity with quite a few details and subtleties''.

Berkman and Vishkin\cite{BV} describe a parallel algorithm for this
problem. Their algorithm can be used to answer queries in constant
time with a single processor (serially). On the Concurrent Read Concurrent
Write model (CRCW), the parallel preprocessing time for the algorithm is
$O(\log^{(m)}n)$, with a near-optimal number of processors, provided
the levels of all nodes and Euler Traversal is given. Without these
assumptions, or on a weaker Concurrent Read Exclusive Write (CREW)
model, the algorithm will take $O(\log n)$ preprocessing time with
nearly optimal number of processors. The sequential implementation of
their algorithm will give a $O(1)$ query time algorithm with nearly
linear preprocessing time.

Menghani and Matani \cite{MM} also describe another simple algorithm.
However, their algorithm takes $O(\log n)$ time to answer queries.

This note describes a very simple $O(1)$ query time algorithm for
finding level ancestors; the preprocessing time is $O(n\log n)$. This
is basically a serial (re)-implementation of the parallel algorithm of
Berkman and Vishkin \cite{BV}.  Ben-Amram \cite{ABA} also gave a
serial version of their parallel algorithm \cite{BV}; however, the
proposed description of the ``basic'' constant-time algorithm is still
simpler and more complete; almost all implementation details are
described.

This implementation of the Berkman-Vishkin algorithm will
offer an alternative to the algorithm of Bender and Farach-Colton.
Students familiar with the Euler-Tour technique \cite{TV} may find this 
conceptually even simpler and, almost certainly, easier to
implement. 
The ancestor of node $v$ at level ``$d$'' is the
first node after $v$ having level $d$ (in the Euler Traversal).
Preprocessing time can be made linear by using table lookup for small
sets. 

The proposed algorithm for table look-up is the usual standard
algorithm. This, or a similar algorithm, has been used, e.g., in
finding the lowest common ancestors\cite{BC:lca}\cite[Section 6.3.1]{LV}
in serial setting and parallel prefix sum problem in parallel
setting\cite{CV}. The table can be constructed in linear serial 
or $O(1)$ parallel time with $n$ processors. As a result, the
parallel algorithm of Berkman and Vishkin\cite{BV} can also be
simplified and made optimal.

All nearest smaller algorithm, which is being used by the algorithms.
is described in Section~2. The preprocessing algorithm is discussed in
Section~3. Answering of queries is discussed in Section~4.  Techniques
for reducing 
preprocessing time are 
discussed in Section~5. 
Table construction is discussed in Section~6.

\section{Preliminaries-Nearest Smallers}

We use two techniques for our algorithm. These are Euler
Traversal\cite{TV,B} and nearest smallers\cite{NS}. The Euler Traversal
Technique is described in Section~3.
Berkman, Schieber and Vishkin \cite{NS} introduced the Nearest
Smallers (NS) problem: %
given an array $A[1:n]$, for each $i$, find the smallest $j>i$ such
that $a_{j}<a_{i}$.

Thus, for each item, we have to find the index of first item (after
it) which is smaller than it.

The nearest smaller problem can be solved serially in linear time using a stack.
The stack will contain indices of all those items whose nearest
smaller has not (yet) been found; thus, ``items in stack'' will be in
increasing order. Stack initially contains ``$1$'', the index of the
first item. The remaining items are picked up one by one. If the item on top
of the stack is larger than the current item, the index of the current
item is the nearest smaller of the stack top. 
Thus, the algorithm to find the nearest smallers for array $A$ is:
\begin{tabbing}
\mbox{top}$=1$; $S[1]=1$ /* stack $S$ contains index of first item */\\
for \= $i=2$ to $n$ do /* look at items one by one */\\
\> while \= (($A[S[\mbox{top}]]>A[i]$)$\&\&$ (top$>0$)) do\\
\>\> $t=S[\mbox{top}]$ /* index at top of stack */\\
\>\> NS$[t]=i$; \mbox{top}$--$ /* Pop item at top of stack */\\
\> \mbox{top}$++$;$S[\mbox{top}]=i$   /* Push current item */\\
\end{tabbing}

As we push an item at most once, the number of pushes is $n$. We can only
pop items which were pushed in the stack. Hence, the number of pop
operations is $O(n)$. Thus, the algorithm takes $O(n)$ time.

The problem can also be solved without using a stack\cite[Lemma~1]{BFN} (see also
\cite{W}).
Consider the following 
algorithm for Nearest Smallers.

Initialise: $A[0]=-\infty$ (default left smaller for items which are
prefix minima). Thus, now each item in array $A[1:n]$ has a left
nearest smaller ``$NS$''.

For first item $NS[1]=0$

for $i=2$ to $n$ do\\
\{
\begin{verse}
$j:=i-1$ /* item on left */\\
while (($A[j]\geq A[i]$) do /* we have not yet found a left smaller */\\
\{\\
\begin{verse}
$j:=NS[j]$ /* As $A[j]$ is larger, $NS[i]$ has to be smaller than
$A[j]$ hence, test $NS[j]$ */
\end{verse}
\}\\
$NS[i]=j$.
\end{verse}
\}

If we assign $j=NS[j]$, then $AS[j]$ is smaller than $A[j]$ thus,
$A[j]$ can not be NS of any item right of $A[i]$.

If we assign $k=NS[i]$, then ``$i$'' will jump to ``$k$'' and ``$j$''
will never be seen again. Hence, assigment $j=NS[j]$ is done only once
for any value of $j$. Or the while look can repeat at most $n$ times,
once for each value of $j$.

\section{Preprocessing} 

We are given a rooted tree, say $T$, which is to be preprocessed to
answer queries of the type: %
find the $k$ th ancestor of a node $v$ in $T$. 

The Euler-traversal technique requires the tree $T$ to be in adjacency
list form. If the initial tree is not in this form, 
we look at
each edge (say) $(u,v)$ in turn and add vertex $u$ to the adjacency list
of $v$ and vertex $v$ to the adjacency list of $u$. Thus, for each
undirected edge $(u,v)$, we are creating two directed edges $(u,v)$
and $(v,u)$. As the in-degree of each node is the same as the
out-degree, the graph is Eulerian, and an Euler tour of the tree can
be found \cite{TV,B} as follows:\\
for each edge $(u,v)$ do
\begin{verse}
The edge after $(u,v)$ in the tour will be the edge $(v,w)$, where $w$ is
the next vertex after $u$ in the adjacency list of $v$.
\end{verse}

We can compute levels of each node (distance from root) in linear
time. The level of the root is zero, and if $w$ is the parent of $v$,
then level$[v]=1+$level$[w]$. Levels can be computed in linear time by
traversing the tree.

Levels can also be computed using Euler Traversal\cite{TV}.
Initially, level is initialised to $0$. When moving from parent to
child, level is incremented, and when going back from child to parent,
level is decremented. If $u$ and $v$ are two successive vertices, then
level$[u]=1\pm $level$[v]$; thus, levels of two successive vertices
differ by exactly one (in absolute terms). 

Let us put the vertices and their levels as they are encountered in
an array (say) $ET$. Thus, the first entry, $ET[0]$ will be
$(\mbox{root},0)$. Each vertex may occur several times. We also store 
the index of any occurrence (say the last) in another array, say
``Position''. Thus, if Position$[v]=i$ then, $ET[i]=(v,\mbox{level}(v))$.

To find the $k$ th ancestor of a node $v$ in $T$, we first find
$i=$Position$[v]$, then we find the first ordered pair (after $i$)
having the first entry as ``level$(v)-k$''. If the ordered pair is
$(\mbox{level}(v)-k,u)$ then $u$ is the $k$ th ancestor of
$v$\cite{BV,ABA}.
Thus, it is sufficient to solve the following Find Smaller (FS)
problem \cite{ABA,BV}:
\begin{quote}
Process an array $A[1:n]$ such that, given query FS$(i,x)$, find the
smallest $j\ge i$ such that $a_{j}\leq x$. 
\end{quote}

The algorithm precomputes the result of some pre-determined queries.
Answering FS-query is then just a table look-up. If the query is
FS$(i,x)$, the numbers $i$ and $x$ are used to determine the location
where the answer (to the query) is available. 

The table, called \cite{BV}, FAR is an array of arrays. 
For each $i$, a different array FAR$_i$ (of size depending on $i$) is
constructed. 
\begin{quote}
The $j$ th entry of the array, FAR$_i[j]$ will contain the index of
the first location right of $a_{i}$ having a value less than or equal
to $a_{i}-j$.
\end{quote}

Thus, FAR$_i[1]$ will contain the index of the first location (after
$i$) with value $a_{i}-1$. For the level ancestor problem, this is the
node with one level less, i.e., it is the parent. FAR$_i[2]$ will
contain the index of the first location with value $a_{i}-2$,
corresponding to the grandparent. In general, FAR$_i[j]$ will contain
the index of the $j$ th ancestor, i.e., if node $v$ is at location $i$
and if $a_i=$level$(v)$, the $d$ th ancestor of $v$ will be at level
$a_i-d$, this is FAR$_{i}[d]$.
FAR-arrays are like storing ancestors for ``ladder'' in the algorithm
of Bender and Farach-Colton\cite{BC}.

Again, from definition, if FAR$_i[j]=h$ then $h$ is the smallest index
(with $h> i$) such that, $a_{h}\leq a_{i}-j$. Given query FS$[i,x]$ we
compute $d=a_{i}-x$. Then FAR$_i[d]$ is the first location right of
$a_{i}$ having value less than or equal to
$a_{i}-d=a_{i}-(a_{i}-x)=x$.

If the depth of the original tree $T$ is $d$, computing all legal
FAR$_i$ values make take $O(nd)$ time (and space) hence only some of
the FAR$_i$ values are computed. 
If $i-1=s2^r$, i.e., $2^r$ is the largest power of $2$ dividing $i-1$,
then we will be computing 
only $3*2^r$ entries in FAR$_i$. For each of the following
numbers: $a_i-1,a_i-2,{\ldots} ,a_i-3*2^r,$ %
we have to find the left most index $k$, $k>i$ such that $a_k\leq
a_i-j$, for $j=1,2,{\ldots} ,3*2^r$. 

All FAR-arrays can be easily computed using Nearest Smallers: Assume
NS$[i]=q$. Then, all items $a_{i+1},{\ldots} ,a_{q-1}$ are larger than
$a_i$. We can make FAR$_i[1]=q$. If $d=a_i-a_q$, then we can also make
FAR$_i[2]=q$,FAR$_i[3]=q$, {\ldots} , FAR$_i[d]=q$. In case, we need
to find FAR$_i[d+1]$, we again find NS$[q]$ and proceed.
Thus, if Nearest Smallers are known, then each FAR entry can be filled
in $O(1)$ time.

\begin{lemma}
All ``FAR'' arrays can be computed in $O(n\log n)$ time and space.
\end{lemma}
Proof If $i-1=s2^r$, i.e., $2^r$ is the largest power of $2$ dividing
$i-1$, then we be computing 
$3*2^r$ entries in FAR$_i$. 

As $\frac{n}{2^i}$ integers are multiple of $2^i$ and
$\frac{n}{2^{i+1}}$ integers are multiple of $2^{i+1}$, it follows
that $\frac{n}{2^i}-\frac{n}{2^{i+1}}=\frac{n}{2^{i+1}}$ integers are
multiple of $2^{i+1}$ but not of $2^i$. Or for $\frac{n}{2^{i+1}}$
integers, the largest power of $2$ which can divide $i-1$ is $2^i$,
hence for these $i$ the size of FAR$_i$ array will be
$3*2^i\left(\frac{n}{2^{i+1}}\right)$, or size of all FAR-arrays
together will be:
$$\sum_{i=1}^{\log n} 3*2^i\left(\frac{n}{2^{i+1}}\right)=\sum_{i=1}^{\log n} \frac{3}{2}n
=\frac{3}{2}n\log n$$
Or computing all FAR-arrays will take $O(n\log n)$ time and space.
$\square$

\section{Query Answering}

For query FS$[i,x]$, $d=a_i-x$ is the difference in depths in the
level ancestor problem. As the depths of two adjacent locations can differ
by at most one, we will not encounter our item if we do $d-1$ places
to the left or to the right. Thus, for these $2(d-1)+1=2d-1$ items,
the value less than or equal to $x$ is the same. Thus, we look at the
index in this range 
divisible by the largest power of two.
The item which we are searching for would have been precomputed and stored.
Details of the method are next described.

Let us assume that the query is FS$[i,x]$. Let $d=a_{i}-x$. 
And let $2^p$ be the largest power of $2$ not larger than $d$, i.e.,
$2^p\leq d<2^{p+1}$. 
For query FS$(i,x)$ , we have to find the first item after location
$i$ smaller than $x$. We proceed as follows:
\begin{enumerate}
\item Let $d=a_i-x$.
\item Let $p$ be s.t., $2^p\leq d<2^{p+1}$, i.e., $2^p$ is the largest
power of $2$ not larger 
than $d$; or $p$ is the number of zeroes in $d$.
\item Let $i_1$ be the largest index less than (or equal to) $i$ s.t.,
$2^p$ divides $i_1-1$, i.e.,
$$i_1=\left\lfloor \frac{i-1}{2^p}\right\rfloor\times 2^p+1$$
\item Return FAR$_{i_1}[a_{i_1}-x]$.
\end{enumerate}

Correctness of the above method follows from:
\begin{lemma}[\cite{BV,ABA}] 
The first element to the right of $a_{i}$ with a value less than or equal to
$x$ is also the first element to the right of $a_{i_1}$ with a value less
than or equal to $x$.
\end{lemma}
Proof As $i-i_1<2^p$, and as $a_{t}$ and $a_{t+1}$ can differ by at
most one, it follows, that for any $i_1\leq j\leq i$,
$a_{j}>a_{i}-2^p>a_{i}-d=x$. $\square$

Thus, we can answer the query by also reporting FS$[i_1,x]$.  And 
FAR$_{i_1}[a_{i_1}-x]$ is the first location right of $a_{i_1}$ with
value less than or equal to $a_{i_1}-(a_{i_1}-x)=x$. As
$a_{i_1}-x<a_i+2^p-x=d+2^p<2^{p+1}+2^p<3*2^p$, value
FAR$_{i_1}[a_{i_1}-x]$ has been computed \cite{BV,ABA}.

\section{Two Level Structure} 

We divide the array (containing levels) into parts of size $k$. For
each part, we find and put the minimum value of that part into another
``global'' array, say $b$. For these $n/k$ minimum values, we
construct a new instance of the FS problem. However, the two adjacent
(minimum) values may now differ by up to $k$. Thus, we have to modify
the algorithm of the previous sections \cite{BV,ABA}.

\subsection{Modified FAR Array}

Let us assume that original array $A$ is divided into $k$ parts and
the minimum item of the $i$ th 
part is kept in location $b_i$ of another array $B$. Now, as two adjacent
items of $B$ can differ (in absolute terms) by $k$, we 
scale
the items of $B$ by dividing each item by $k$ and keeping only the
integer part. As a result, as one ``scaled'' item differs from the next
item by at most one, we can use the algorithm of the previous section.
However, instead of returning the next node
with value $d$ (when the query is for $d/k$), the query will return 
the next node with value $k\left\lfloor \frac{d}{k}\right\rfloor$. 
Thus, we are getting an index with ``coarse'' or approximate location
of the cell with the value of $d$. To get ``fine'' or the exact
index of the location 
with the value $d$, another array of size $k$ is used. A more detailed
description is given next. 

We will call the first array of this section the ``modified'' mod\_FAR
array to differentiate it from the FAR array of Section~3; this is
actually the FAR array as defined in \cite{BV}.

Let $e=\left\lfloor \frac{b_i}{k}\right\rfloor$ (thus,
$(e-1)k<b_{i}\leq ek$). Again, we let $2^r$ be the largest power of
$2$ which divides $i-1$. Value mod\_FAR$_{i}[j]$ will give the first
location right of $A[b_{i}]$ with value less than or equal to $(e-j)k$,
again for $1\leq j\leq 3*2^r$. 

Let $i_1$ be as before. And let $e_1=\left\lfloor
\frac{b_{i_1}-x}{k}\right\rfloor$. Then mod\_FAR$_{i_1}[e_1]$ will give the
first location right of $A[b_{i_1}]$ with value less than or equal to
$(e-e_1)k=k\left( \left\lfloor \frac{b_{i_1}}{k}\right\rfloor-
\left\lfloor \frac{b_{i_1}-x}{k}\right\rfloor\right)\leq
k\left(1+\left\lfloor \frac{x}{k}\right\rfloor\right)\leq k+x$.

For computing the 
mod\_FAR-values, we use another array $C[1:n]$, with $c_i=\lfloor
\frac{b_i}{k}\rfloor$. And find the nearest smallers in the $C$-array
(assuming that in case of duplicates, the first entry is smaller).
Now, $e=\left\lfloor \frac{b_i}{k}\right\rfloor=c_i$. If
mod\_FAR$_{i}[j]=t$, then $b_t$ is the first number right of $b_{i}$
in $A$ with value less than or equal to $(e-j)k=(c_i-j)k$; or
equivalently, $b_t/k$ is is the first number right of $b_{i}$ in $B$
with value less than or equal to $(c_i-j)$. Thus, the 
mod\_FAR-values can be computed as in Section~3, using array $C$
instead (of $A$).

Remark To make sure that all divisions are by a power of $2$, we
can choose $k$ to be a number between $\frac18 \log n$ and $\frac14
\log n$ which is a power of $2$. 

From previous analysis, we know that the number of entries of mod\_FAR
will be $O(n'\log n)=O(\frac{n}{k}\log n)$ If $k=\theta(\log n)$, this
will be $O(n)$.
Hence, we can preprocess the global array in $O(\frac{n}{k}\log
n+\frac{n}{k}k)=O(n)$ time, if $k=\theta(\log n)$.
Processing of local parts is discussed in Section~6 and in the appendix.

\subsection{Near Array}

Using the mod\_FAR array, we were able to get ``coarse'' location of
the desired item.
To get the ``exact'' location, we use another array Near \cite{BV}.
Near$_i[j]$, for $1\leq j\leq k$ will give the first location right of
$b_{i}$ with value less than $b_{i}-j$ (just like the FAR array of
Section~3). Note that we are storing $k$ entries for each $b_{i}$.
Thus, the total number of entries will be
$O\left(\frac{n}{k}k\right)=O(n)$. If $d=b_{i}-x$.  Then Near$_i[d]$
is the first location right of $b_{i}$ having value less than or equal
to $b_{i}-d=b_{i}-(b_{i}-x)=x$ (provided, $d\leq k$).

Again, the Near-table can be filled using Nearest smallers value, with
$O(1)$ time per entry.

\subsection{Query}

We next describe the query algorithm on the array $B$. This is more or
less the algorithm of Section~4.

Let us assume that the query is FS$[i,x]$. Let $d=b_{i}-x$. 
And let $2^p$ be the largest power of $2$ not larger than $d$, i.e.,
$2^p\leq d<2^{p+1}$. 
For query FS$(i,x)$ , we have to find the first item, in array $B$,
after location $i$ smaller than $x$. We proceed as follows:
\begin{enumerate}
\item Let $d=b_i-x$.
\item Let $p$ be s.t., $2^p\leq d<2^{p+1}$, i.e., $2^p$ is the largest
power of $2$ not larger 
than $d$. 
\item Let $i_1$ be the largest index less than (or equal to) $i$ s.t.,
$2^p$ divides $i_1-1$, i.e.,
$$i_1=\left\lfloor \frac{i-1}{2^p}\right\rfloor\times 2^p+1$$
\item If $b_{i_1}-x<k$, then return Near$_{i_1}[b_{i_1}-x]$
else let $i_2=$mod\_FAR$_{i_1}[b_{i_1}-x]$.
\item return Near$_{i_2}[b_{i_2}-x]$
\end{enumerate}

Again we have:
\begin{lemma}[\cite{BV,ABA}] 
First element to the right of $b_{i}$ with value less than or equal to
$x$ is also the first element to the right of $b_{i_1}$ with value less
than or equal to $x$.
\end{lemma}
Proof As $i-i_1<2^p$, and as $b_{t}$ and $b_{t+1}$ can differ by at
most $k$, it follows, that for any $i_1\leq j\leq i$,
$b_{j}>b_{i}-k2^p>b_{i}-d=x$. $\square$

Thus, we can answer the query by also reporting FS$[i_1,x]$.
As $0<b_{i_1}-x<3k*2^p$, we have $0\leq
\frac{b_{i_1}}{k}-\frac{x}{k}<3*2^p$, or $0\leq
c_i-\frac{x}{k}<3*2^p$.

\section{Table Construction and Putting Everything Together}

For queries to the original FS problem, we first determine the index
of the ``local'' group (using the global algorithm). Then, we have
to determine the index in the corresponding local group (of
$k$-items).

As in the FS problem, we are to return the first location for each
group. Thus, it is sufficient to store the first occurrence of each value in
the group (i.e., the first occurrence of each level in the original
problem). As values in a group increase or decrease by one, there can
be at most $k$ different values in a group. 
Thus, using an array of size $k$ for each group is enough.
A single scan can easily fill these arrays in $O(k)$ time. 
In parallel, on the Priority-CRCW model,
the arrays can be filled in $O(1)$ time with $k$ processors.

However, it is quite possible that the answer to FS$(i,x)$ lies in the
same group containing $i$. Thus, we need to preprocess each group to
answer queries (assuming the answer is in that group). We can do
this in two different ways
\begin{itemize}
\item Process each group for FS$(i,x)$ queries using the algorithm of
Section~3. This approach is discussed in the Appendix.

\item Create a table and do a table lookup. This is discussed in the rest
of this section.
\end{itemize}

Let us choose $k$ to be a power of $2$ between $\frac18 \log n$ and
$\frac14 \log n$. As adjacent entries (or levels) differ by plus or
minus one, we can interpret a string of $k$ pluses or minuses as a
binary string (say $1$ for plus and $0$ for minus). There are $2^k$
such strings. Each string (or row in the table) can be interpreted as a
hypothetical input\cite{CV,BC:lca}. 

We process each  string   of length $k$ independently. As there are $k$
entries   in each  string (or  hypothetical group), the largest legal index
can be $k$ for FAR-values. Using the  algorithm  of Section~3, for each
item $i$  in the  hypothetical group, FAR$_i[1:k]$  can be computed  in
$O(k)$  time. Thus, we spend $O(k^2)$  time  per group.   As there  are
$2^k$ hypothetical   groups, total   time will be $O(k^22^k)=O(n)$   if
$k\leq (1-\epsilon)\log n$, for any $\epsilon>0$.

Thus, we have the following lemma:
\begin{lemma}
There is an algorithm to solve FS-problem, in which two items differ
by at most one with $O(n)$ preprocessing time and $O(1)$ query time. 
\end{lemma}

In the parallel setting, we process each row of the table or each
hypothetical input in parallel using the algorithm of Berkman and
Vishkin\cite[Section 3.1]{BV}. The parallel algorithm for each
hypothetical input takes $O(1)$ time with $k\log^3 k$ processors (see
remarks after Step~1 in \cite{BV}); actually, we can also use the
``simpler'' 
$n^2$ processor, $O(1)$ time algorithm for finding the minimum. As each
hypothetical input is processed independently and in parallel, the total
time for preprocessing is $O(1)$. Next $k$ processors are assigned to
each hypothetical input; these query for at most $k$ legal values
(levels of ancestors); as query time is $O(1)$, the total time will remain
$O(1)$. Number of processors used will, by analysis similar to that of
the serial case, will be $O(n)$, for $k\leq (1-\epsilon)\log n$, for
any $\epsilon>0$.

For each group, we have to store a pointer to the corresponding row in
the table. Which can be done in linear time in a sequential setting. In
the parallel setting, we can proceed as in \cite{CV} (see precomputation
routine) if we do not have an instruction to ``pack'' a string of $k$
bits into a single word.
If we are to use $n$ processors, 
we should choose 
$k=O(\log\log n)$.

Thus using the algorithm of Section~3.1 of \cite{CV} together with the
proposed table construction and lookup routines, we get a parallel
algorithm with $O(1)$ preprocessing time with $n$ processors. Queries
can be answered in $O(1)$ time with a single processor.

\subsection*{\center{Acknowledgements}}

I wish to thank students who attended lectures of CS602 (2023-2024) for their comments
and reactions on a previous version. 

\bibliography{general}

\begin{thebibliography}{99}
\bibitem{AH}
S.Alstrup and J.Holm, Improved Algorithms for Finding Level
Ancestors in Dynamic Trees. ICALP 2000: 73-84 (2000)

\bibitem{BFN}
J.Barbay, J.Fischer abd G.Navarro,
LRM-Trees: Compressed indices, adaptive sorting, and compressed
permutations, In: Giancarlo, R., Manzini, G. (eds) Combinatorial Pattern Matching. CPM 2011. Lecture Notes in Computer Science, vol 6661. Springer, Berlin, Heidelberg. 
(also arXiv:1009.5863) 

\bibitem{ABA} A.M. Ben-Amram, The Euler Path to Static
Level-Ancestors. CoRR abs/0909.1030 (2009)

\bibitem{B}B.G.Baumgart, A
polyhedron representation for computer vision, Proc. 1975 National
computer conf., AFIPS conference proceedings vol 44, 589-596 (1975).

\bibitem{BC}
M.A.Bender and M.Farach-Colton, The Level Ancestor
Problem Simplified. Theor. Comput. Sci. 321: 5-12(2004).

\bibitem{BC:lca} M.A.Bender and M.Farach-Colton,  The LCA
Problem Revisited. In: Gonnet, G.H., Viola, A. (eds) Theoretical
Informatics. LATIN 2000. LNCS 1776 (2000) Springer, Berlin, Heidelberg.

\bibitem{BV} O. Berkman and U. Vishkin. Finding level-ancestors in
trees. J. of Comp. and Sys. Scie., 48(2):214-230 (1994).

\bibitem{NS}
O.Berkman, B.Schieber and U.Vishkin: Optimal Doubly Logarithmic
Parallel Algorithms Based on Finding All Nearest Smaller Values. J.
Algorithms 14(3): 344-370 (1993)

\bibitem{CV} R.Cole and U.Vishkin, Faster optimal parallel prefix sums
and list ranking. Information and Computation  81 (3):
334--352 (1989).

\bibitem{D}P. Dietz. Finding level-ancestors in dynamic trees. In 2nd Work. on
Algo. and Data Struc., LNCS 1097, 32-40 (1991).

\bibitem{LV} G.M.Landau and U.Vishkin, Chapter 6: Approximate String
Searching, in Pattern Matching
Algorithms Ed. Alberto Apostolico and Zvi Galil, Oxford University Press,
1997.

\bibitem{MCPP} M.Mabrey, T.Caputi, G.Papamichail, D.Papamichail,
	Static Level Ancestors in Practice.  CoRR abs/1402.2741 (2021)

\bibitem{MM} G.Menghani, D.Matani, A Simple Solution to the
Level-Ancestor Problem, arXiv:1903.01387v2 (2021).

\bibitem{TV}R.E.Tarjan and U.Vishkin, An Efficient Parallel
Biconnectivity Algorithm. SIAM J. Comput. 14(4): 862-874 (1985)

\bibitem{W} All Nearest Smaller Values, Wikipedia, Id=1188208021


\end{thebibliography}

\begin{appendices}

\section{Multilevel Structure}

Each local group of $k$ items can be preprocessed using the algorithm
of Section~3, to answer intra-group queries. Time to preprocess each
group will be $O(k\log k)$. As there are $n/k$ groups, the total
preprocessing time will be $O(\frac{n}{k}(k\log k)=O(n\log\log n)$, if
$k=O(\log n)$; query time is still $O(1)$. Thus,
\begin{lemma}\label{double}
There is an algorithm to solve FS-problem, in which two items differ
by at most one with $O(n\log\log n)$ preprocessing time and $O(1)$
query time. 
\end{lemma}

Instead of using the basic algorithm of Section~3, for local groups,
we can use the algorithm of Lemma~\ref{double} instead. We, as before,
divide the array (containing levels) into parts of size $k=\theta(\log
n)$. But for each part, we preprocess each group using the algorithm
of Lemma~\ref{double} in $O(k\log\log k)$ time. Or total time for
preprocessing all groups is
$O\left(\frac{n}{k}k\log^{(3)}n\right)=O(n\log^{(3)}n)$.

By using a constant number of levels, the preprocessing time can be
made $O(n\log^{(r)}n)$, for any $r>1$. Query time will be $O(r)$.

Remark 
For most practical values of $n$, a two or
three-level structure is likely to be enough 
(as $\log^{(3)} N\leq 3$, for $N<10^{75}$).
\end{appendices}

\end{document}